\documentclass{article}
\usepackage{spconf}

\usepackage[utf8]{inputenc} 
\usepackage[T1]{fontenc}    
\usepackage{hyperref}       
\usepackage{url}            
\usepackage{booktabs}       
\usepackage{amsfonts}       
\usepackage{nicefrac}       
\usepackage{microtype}      
\usepackage{xcolor}         
\usepackage{kotex}
\usepackage{amsmath}
\usepackage{amssymb}
\usepackage{graphicx}
\usepackage{subcaption}
\usepackage{multirow}
\usepackage{makecell}       
\usepackage{pifont}         

\newcommand{\tblcheck}{\textcolor{green!55!black}{\ding{51}}}
\newcommand{\tblcross}{\textcolor{red!75!black}{\ding{55}}}

\ninept
\begin{document}
\title{StreamSE: Neural Audio Codec-based Streaming Speech Enhancement for On-device Inference}

\name{Hyunjin Choi, Zinu Kim, Junwon Moon, Jeongyun Chae, Seungbeom Kim, Kyuhong Shim}
\address{Sungkyunkwan University, Republic of Korea \\
\texttt{\small emily.choi@skku.edu, khshim@skku.edu}
}

\maketitle

\begin{abstract}
Recent generative speech enhancement (SE) models achieve high perceptual quality, but their non-causal processing and expensive computation make them difficult to use for streaming, on-device inference.
We propose StreamSE, a 20M-parameter streaming SE model that operates in the latent space of a frozen pretrained neural audio codec.
At each frame, StreamSE directly regresses a clean continuous latent from the noisy continuous latent while conditioning on its own predictions from previous frames.
Enhancement therefore requires a single forward pass per frame and no iterative sampling or look-ahead.
We further truncate the codec's residual vector quantizer (RVQ) to its coarse levels, halving the number of tokens per frame while improving robustness to reverberation.
StreamSE is the only streaming generative model in our comparison, yet it matches or exceeds the perceptual quality of the offline generative baselines.
Furthermore, the complete StreamSE pipeline runs in real time on a single smartphone CPU core.
Our results show that continuous latent regression provides a practical route to generative SE under fully causal, on-device streaming constraints.
\end{abstract}


\section{Introduction}\label{sec:intro}
Speech enhancement recovers clean speech from noisy signals and is widely used to improve speech quality in voice communication and to preprocess input for speech recognition.
Applications such as telephony, hearing aids, and on-device voice assistants impose two constraints simultaneously: causality requires that the output depend only on current and past frames, while the computational budget requires that each frame be processed within one frame on device.

Deep learning-based speech enhancement has progressed from discriminative methods that map noisy speech to clean speech to generative methods that employ diffusion models and language models.
In particular, SELM~\cite{wang2024selm} generates clean speech as discrete tokens with a language model, achieving higher perceptual quality than discriminative methods.
However, existing generative approaches typically rely on non-causal architectures. 
SELM, for instance, processes an entire utterance at once.
Diffusion-based methods further incur the computational cost of iterative sampling at inference~\cite{lu2022conditional}.
Attaining high perceptual quality under both causal streaming and on-device computational constraints, therefore, remains challenging.

\begin{figure*}[t]
     \centering\includegraphics[width=0.88\linewidth]{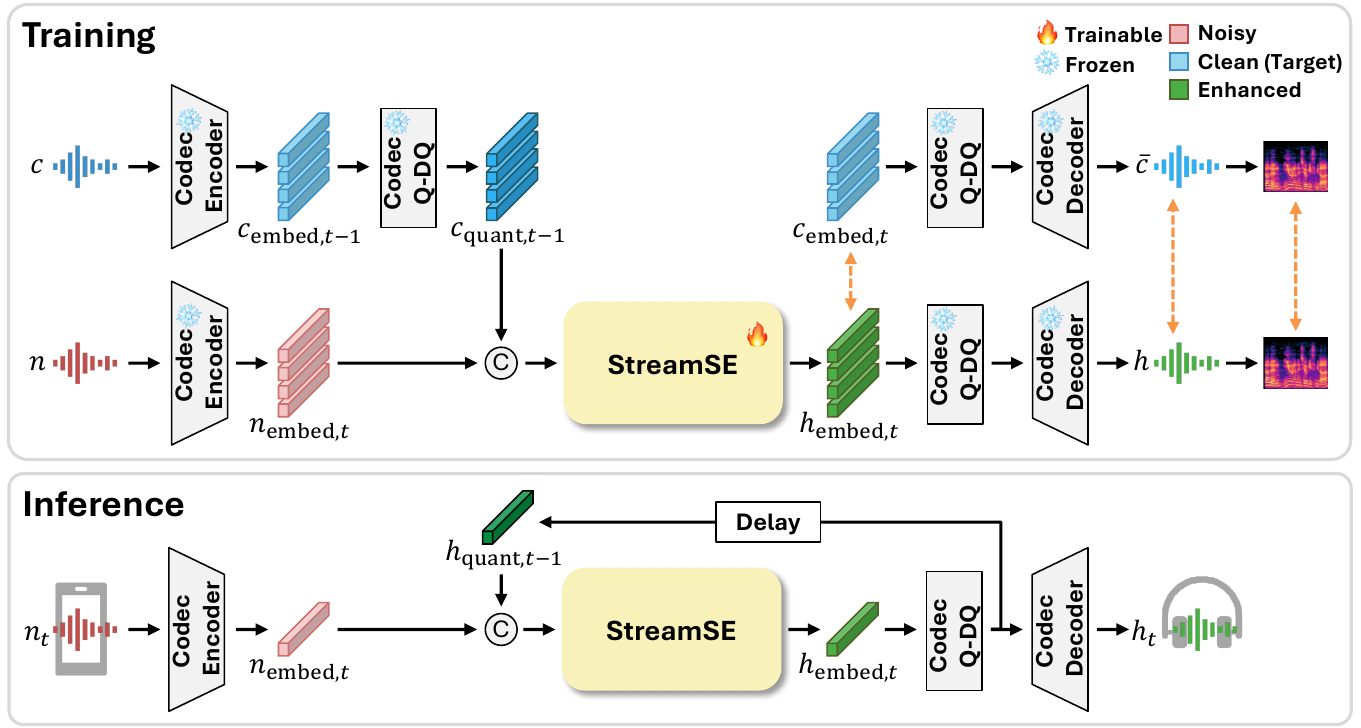}
     \caption{Overview of StreamSE during training (top) and inference (bottom).
     \textcircled{\scriptsize C} denotes concatenation, and Q-DQ denotes $\mathrm{QDQ}_k$, quantization followed by dequantization, which maps a continuous latent onto the sum of its selected RVQ codebook vectors.
     Dashed orange arrows mark the training objective, which compares the predicted latent against the clean latent $c_{\mathrm{embed}}$ and the decoded waveform and its mel spectrogram against the codec round-trip reference $\bar{c}$.
     At inference, the delay block feeds the previous quantized-and-dequantized enhanced latent back as conditioning; the figure draws a single previous frame, whereas the model attends to all past frames.}
     \label{fig:method}
\end{figure*}

To address this problem, we propose StreamSE, a lightweight streaming speech enhancement model that operates in the latent space of a frozen pretrained neural audio codec (Fig.~\ref{fig:method}).
The entire pipeline, including the codec encoder and decoder, is fully causal.
At each frame, the codec encoder maps noisy speech to a continuous latent, and StreamSE directly regresses the corresponding clean latent.
The codec quantizer and decoder then reconstruct the waveform from the resulting latent.
To maintain temporal consistency, the model feeds its own quantized predictions back as conditioning for subsequent frames.
By predicting a continuous latent rather than discrete codec tokens, StreamSE avoids separately predicting a residual vector quantizer (RVQ) token at each level and completes enhancement with a single forward pass per frame.
We additionally apply level truncation, which retains only the eight coarse levels of the RVQ and halves the number of tokens per frame, discarding fine levels that we expect to contain residual reverberation.

On the DNS Challenge 2020 test set, StreamSE achieves the best score on 7 of the 9 DNSMOS metrics, including all 6 metrics in the two synthetic categories, despite being the only streaming generative model among the compared systems.
Measurements on a single CPU core of a Galaxy S22 mobile device further show that the full int8-quantized pipeline operates in real time, consuming 76\% of the 80\,ms per-frame budget, with StreamSE itself requiring only 5.46\,ms.
\section{Related Work}\label{sec:related}

Deep learning-based speech enhancement has traditionally relied on discriminative methods that map noisy speech to clean speech in the time domain~\cite{luo2019conv} or the time-frequency domain~\cite{chen2023inter}.
These methods trade noise suppression against signal preservation under severe degradation.
Generative methods instead regenerate clean speech rather than masking or filtering the noisy input, either with diffusion or flow matching conditioned on the degraded input~\cite{lu2022conditional, wang2025flowse, lemercier2023storm} or with language models that generate clean discrete tokens~\cite{wang2024selm, yao2025gense, yang2024genhancer}.
They often outperform discriminative baselines on perceptual metrics, but at the cost of non-causal processing or iterative sampling.

A growing body of work uses a neural audio codec as the representation and generates clean codec tokens conditioned on the degraded input~\cite{yang2024genhancer}, sometimes through an intermediate semantic stage~\cite{yao2025gense}.
Recent studies report that predicting continuous latents often outperforms predicting discrete tokens, and that vector quantization itself contributes to robustness regardless of the prediction target~\cite{kammoun2026modeling}.
Work in this direction enhances the pre-quantization embeddings of a frozen codec with an embedding-level loss~\cite{li2025speech}, but it is evaluated only in a non-streaming, utterance-level setting and targets cloud deployment.

Streaming speech enhancement is predominantly discriminative and typically uses a short look-ahead~\cite{defossez20_interspeech}, and autoregressive conditioning on a model's own past output has been studied at low latency~\cite{andreev23_interspeech} and in multichannel settings~\cite{shen25_interspeech}.
Token-based generative approaches remain largely unexplored under streaming constraints, and the reported work relies on a codec specifically designed for low latency~\cite{xue2024low}.
In addition, the cost of these approaches is usually reported as a real-time factor on a server GPU rather than measured per frame on the target device.
Our work differs in two respects: (1) it combines continuous latent prediction with fully causal streaming that requires no look-ahead, and (2) it measures per-frame runtime directly on the target device.
\section{Method}\label{sec:method}

\begin{table*}[t]
    \centering
    \caption{Comparison of different systems on the DNS Challenge test set. Streamable indicates whether the evaluated configuration supports streaming inference (\tblcheck\ yes, \tblcross\ no). StreamSE rows differ in the number of look-ahead frames $\ell$; the algorithmic latency is $(1{+}\ell)$ frames of 80\,ms. The int8 row is the $\ell{=}0$ model run through the quantized on-device pipeline of Section~\ref{ssec:experiment_ondevice}.}
    \label{tab:results}
    \footnotesize
    \setlength{\tabcolsep}{4pt}
    \renewcommand{\arraystretch}{1.05}
    \resizebox{1.0\linewidth}{!}{
    \begin{tabular*}{\linewidth}{@{\hspace{2\tabcolsep}\extracolsep{\fill}} l c ccc ccc ccc @{\hspace{2\tabcolsep}}}
        \toprule
        \multirow{3}{*}{System} & \multirow{3}{*}{Streamable} & \multicolumn{3}{c}{With Reverb} & \multicolumn{3}{c}{Without Reverb} & \multicolumn{3}{c}{Real Recordings} \\
        \cmidrule(lr){3-5} \cmidrule(lr){6-8} \cmidrule(lr){9-11}
         & & \multicolumn{3}{c}{DNSMOS $\uparrow$} & \multicolumn{3}{c}{DNSMOS $\uparrow$} & \multicolumn{3}{c}{DNSMOS $\uparrow$} \\
        \cmidrule(lr){3-5} \cmidrule(lr){6-8} \cmidrule(lr){9-11}
         & & SIG & BAK & OVL & SIG & BAK & OVL & SIG & BAK & OVL \\
        \midrule
        Noisy & -- & 1.760 & 1.497 & 1.392 & 3.392 & 2.618 & 2.483 & 3.053 & 2.510 & 2.255 \\
        \midrule
        Conv-TasNet~\cite{luo2019conv} & \tblcross & 2.415 & 2.710 & 2.010 & 3.092 & 3.341 & 3.001 & 3.102 & 2.975 & 2.410 \\
        Demucs~\cite{defossez20_interspeech} & \tblcheck & 2.510 & 2.641 & 2.215 & 3.124 & 3.257 & 3.010 & 2.974 & 2.870 & 2.291 \\
        Inter-SubNet~\cite{chen2023inter} & \tblcross & 2.651 & 2.581 & 2.362 & 3.458 & 3.821 & 3.099 & 3.258 & 3.574 & 2.806 \\
        \midrule
        CDiffuSE~\cite{lu2022conditional} & \tblcross & 2.541 & 2.300 & 2.190 & 3.294 & 3.641 & 3.047 & 3.201 & 3.104 & 2.781 \\
        SGMSE~\cite{welker2022speech} & \tblcross & 2.730 & 2.741 & 2.430 & 3.501 & 3.710 & 3.137 & 3.297 & 2.894 & 2.793 \\
        StoRM~\cite{lemercier2023storm} & \tblcross & 2.947 & 3.141 & 2.516 & 3.514 & 3.941 & 3.205 & 3.410 & 3.379 & 2.940 \\
        SELM~\cite{wang2024selm} & \tblcross & 3.160 & 3.577 & 2.695 & 3.508 & 4.096 & 3.258 & \textbf{3.591} & 3.435 & \textbf{3.124} \\
        \midrule
        \textbf{StreamSE ($\ell{=}2$)}  & \tblcheck & \textbf{3.505} & \textbf{4.054} & \textbf{3.207} & \textbf{3.630} &         4.138 &         3.380 & 3.432 & \textbf{3.879} & 3.075 \\
        \textbf{StreamSE ($\ell{=}1$)}   & \tblcheck &         3.493 &         4.044 &         3.189 &         3.628 & \textbf{4.143} & \textbf{3.382} & 3.410 &         3.875 & 3.054 \\
        \textbf{StreamSE ($\ell{=}0$)}   & \tblcheck &         3.417 &         3.990 &         3.086 &         3.617 &         4.124 &         3.361 & 3.399 &         3.830 & 3.016 \\
        \midrule
        \textbf{StreamSE ($\ell{=}0$, int8)} & \tblcheck &         3.356 &         3.867 &         2.987 &         3.586 &         4.088 &         3.315 &         3.327 &         3.707 &         2.914 \\
        \bottomrule
    \end{tabular*}}
\end{table*}

\subsection{Architecture Overview}\label{ssec:method_overall}
Given noisy speech $n$, streaming speech enhancement recovers clean speech $c$ under a causal constraint: the output at time $t$ must depend only on frames $1,\dots,t$ of the input.
The proposed architecture comprises the encoder, the RVQ, and the decoder of a pretrained neural audio codec, together with a lightweight speech enhancement module, StreamSE, placed between the encoder and the RVQ.
Only StreamSE is trained; the codec remains frozen throughout.

The full inference pipeline follows Fig.~\ref{fig:method} and Eq.~\eqref{eq:se}: the codec encoder maps the noisy waveform $n$ to continuous latents $n_{\mathrm{embed}}$, StreamSE predicts the enhanced latents $h_{\mathrm{embed}}$, and the codec decoder reconstructs the enhanced waveform $h$ from the quantized prediction $h_{\mathrm{quant}}$.
At time $t$, the pipeline is
\begin{equation}
    \begin{aligned}
        n_{\mathrm{embed},t} &= \mathrm{Encoder}(n)_{t} , \\
        h_{\mathrm{embed},t} &= \mathrm{StreamSE}\!\left(n_{\mathrm{embed},\le t},\; h_{\mathrm{quant},<t}\right) , \\
        h_{\mathrm{quant},t} &= \mathrm{QDQ}_{k}(h_{\mathrm{embed},t}) , \\
        h_{t} &= \mathrm{Decoder}(h_{\mathrm{quant},\le t})_{t} .
    \end{aligned}
    \label{eq:se}
\end{equation}
Here, $n_{\mathrm{embed}}, h_{\mathrm{embed}}\in\mathbb{R}^{T\times d}$ denote the noisy latent and the enhanced latent, respectively, where $T$ is the number of frames and $d$ is the latent dimension per frame.
$\mathrm{QDQ}_{k}$ quantizes a latent into discrete tokens using the first $k$ coarse RVQ levels and then dequantizes them into a continuous latent as the sum of the corresponding codebook vectors.
For the clean speech $c$, we similarly define
\begin{equation}
    \begin{aligned}
        c_{\mathrm{embed}} &= \mathrm{Encoder}(c) , \\
        c_{\mathrm{quant}} &= \mathrm{QDQ}_{k}(c_{\mathrm{embed}}) .
    \end{aligned}
    \label{eq:clean}
\end{equation}
Within StreamSE the representation remains continuous; each frame is quantized once, at the output, and this quantized latent serves both decoding and the context of the next frame.

\subsection{Neural Audio Codec}\label{ssec:method_codec}
We employ MOSS-Audio-Tokenizer-Nano~\cite{gong2026moss} as the neural audio codec.
It is a causal streaming codec with approximately 22M parameters, including the quantizer, and processes a 48\,kHz waveform at 12.5\,Hz.
Each 80-ms frame is encoded into a $d{=}768$-dimensional continuous latent and quantized using an $L{=}16$-level RVQ with a codebook size of 1,024.
We select this codec because it is causal and operates at a low frame rate.
Because the codec is trained with quantizer dropout~\cite{gong2026moss}, its decoder reconstructs the waveform from only a prefix of the RVQ levels.
We exploit this property and truncate the RVQ to its $k$ coarse levels, applying $\mathrm{QDQ}_{k}$ with the same $k$ in the conditioning path, the auxiliary losses, and the decoding path.

\subsection{StreamSE}\label{ssec:method_se}
StreamSE is a decoder-only causal Transformer with approximately 20M trainable parameters, excluding the frozen codebook embeddings.
The module is autoregressive across frames but directly regresses a continuous latent before quantization instead of classifying discrete tokens.
The input at time $t$ is the concatenation of the noisy latent $n_{\mathrm{embed},t}$ and the quantized latent of the previous frame, $c_{\mathrm{quant},t-1}$ during training and $h_{\mathrm{quant},t-1}$ at inference; earlier frames enter through causal self-attention, and the context of the first frame is a zero vector.
During training, we apply teacher forcing by feeding the ground-truth quantized latents $c_{\mathrm{quant},<t}$ as the previous-frame context and process the entire sequence in parallel.
At inference, the module operates autoregressively and feeds its own quantized past predictions $h_{\mathrm{quant},<t}$ back as context (Eq.~\eqref{eq:se}), so that each prediction is conditioned on the enhanced past and consecutive frames stay temporally consistent.

Unlike our continuous regression, autoregressive classification over discrete RVQ tokens requires repeated per-level classification, with computational cost increasing with the number of RVQ levels.
This overhead would be undesirable under the tight per-frame budget of on-device streaming inference.
StreamSE leaves high-fidelity waveform reconstruction to the pretrained codec and converts the enhancement task to predicting a single continuous latent per frame.

\subsection{Loss Function}\label{ssec:method_loss}
The training objective consists of a regression loss in the latent space and two auxiliary losses computed along the decoding path.
The regression loss $\mathcal{L}_{\text{emb}}$ is the L1 distance between the predicted latent $h_{\mathrm{embed}}$ and the clean latent $c_{\mathrm{embed}}$.
The two auxiliary losses $\mathcal{L}_{\text{wav}}$ and $\mathcal{L}_{\text{mel}}$ compare two waveforms produced by the same quantization--decoding path: the enhanced output $h = \mathrm{Decoder}(h_{\mathrm{quant}})$ and the codec round-trip reference $\bar{c} = \mathrm{Decoder}(c_{\mathrm{quant}})$.
$\mathcal{L}_{\text{wav}}$ compares them in the waveform domain, and $\mathcal{L}_{\text{mel}}$ compares them in the log-mel spectrogram ($\mathrm{LogMel}$) domain.
Using the codec reconstruction $\bar{c}$ rather than the original clean waveform $c$ removes the reconstruction error of the frozen codec from the auxiliary training signal, so the losses focus on the prediction error of StreamSE:
\begin{equation}
    \begin{aligned}
        \mathcal{L}_{\text{emb}} &= \lVert h_{\mathrm{embed}} - c_{\mathrm{embed}} \rVert_1 , \\
        \mathcal{L}_{\text{wav}} &= \lVert h - \bar{c} \rVert_1 , \\
        \mathcal{L}_{\text{mel}} &= \lVert \mathrm{LogMel}(h) - \mathrm{LogMel}(\bar{c}) \rVert_1 .
    \end{aligned}
    \label{eq:losses}
\end{equation}
The total loss is the weighted sum of the three terms above, with $\lambda_{w}{=}0.2$ and $\lambda_{m}{=}0.5$:
\begin{equation}
    \mathcal{L} = \mathcal{L}_{\text{emb}}
        + \lambda_{w}\,\mathcal{L}_{\text{wav}}
        + \lambda_{m}\,\mathcal{L}_{\text{mel}}.
\end{equation}
Following FlowSE~\cite{wang2025flowse}, we compute the mel loss using 100-bin log-mel spectrograms with a hop length of 512 samples (10.67 ms) and a window length of 2,048 samples (42.67 ms) for a 48 kHz audio.
To train the model end-to-end, we use a straight-through estimator (STE) to propagate gradients through quantization while keeping the codec parameters frozen.
\section{Experiments}\label{sec:experiment}

\subsection{Datasets and Evaluation Metrics}\label{ssec:experiment_data}

\textbf{Training Data.}
We synthesize training pairs online from the clean speech of LibriTTS-R~\cite{koizumi2023libritts} train-clean-100 and train-clean-360 and of the VoiceBank-DEMAND training set~\cite{valentini2016investigating}.
For each pair, we mix noise from WHAM!~\cite{wichern2019wham} or DEMAND~\cite{thiemann2013demand} with probability 0.8 at a signal-to-noise ratio (SNR) drawn from $-5$ to $20$\,dB.
We add reverberation with probability 0.5 from the openSLR26 and openSLR28 room impulse responses (RIRs)~\cite{ko2017study}.
We rescale both signals by a common gain such that the noisy signal has an RMS sampled uniformly from $-35$ to $-15$\,dBFS.
We also band-limit both the clean and noisy signals by resampling to 8, 16, or 24\,kHz and back to 48\,kHz for regularization.
For pairs with added reverberation, the target is the dry signal, so the model learns to remove reverberation rather than to reproduce it.

\vspace{0.2cm}\noindent\textbf{Test Data and Evaluation Metrics.}
We evaluate on the three categories of the DNS Challenge 2020~\cite{reddy2020interspeech} test set.
Because generative-model outputs are not necessarily sample-aligned with their references, we use the reference-free DNSMOS P.835 (SIG/BAK/OVL).
For the level-truncation ablation, we additionally report speaker similarity (SIM), the cosine similarity of WeSpeaker~\cite{wang2023wespeaker} embeddings, rescaled to $[0,1]$ as $(\cos+1)/2$.

\subsection{Training Setup and Baselines}\label{ssec:experiment_setup}
StreamSE is a causal Transformer with 512-dimensional hidden representations, 6 layers, and 8 attention heads.
It uses rotary position embeddings (RoPE).
At inference, RoPE uses the current frame index as the position offset. 
No causal mask is needed because each query attends only to cached past keys and the current key.
We set the number of coarse RVQ levels to $k{=}8$ as default.
We apply dropout rate of 0.2 and weight decay of 0.1, and optimize with AdamW.
The learning rate warms up linearly to a peak of $6{\times}10^{-4}$ over the first 1{,}000 steps and then decays linearly until the end of training.
We clip gradients to unit norm.
Training uses segments of 125 codec frames, that is, 10\,s, matching the KV cache width used at inference.
We train for 20 epochs on 16\,GB V100 GPU with an effective batch size of 48.

The compared systems in Table~\ref{tab:results} fall into two families.
The discriminative family comprises Conv-TasNet~\cite{luo2019conv}, Demucs~\cite{defossez20_interspeech}, and Inter-SubNet~\cite{chen2023inter}, while the generative family comprises the diffusion-based CDiffuSE~\cite{lu2022conditional}, SGMSE~\cite{welker2022speech}, and StoRM~\cite{lemercier2023storm} together with the token-based language model SELM~\cite{wang2024selm}.
All baseline numbers are quoted from the SELM paper.

\subsection{Results}\label{ssec:experiment_main}
Table~\ref{tab:results} reports the comparison on the DNS Challenge test set.
Each fp32 StreamSE configuration outperforms all external baselines on seven of the nine DNSMOS metric–category combinations, even though StreamSE is the only causal, streamable architecture among the compared generative systems.
Unless stated otherwise, we discuss the $\ell{=}0$ configuration, which uses no look-ahead.
This configuration is the one measured on device in Section~\ref{ssec:experiment_ondevice} and shares its checkpoint with the $k{=}8$ row of Table~\ref{tab:ktrunc}.
The particularly strong performance under reverberation is examined in the level-truncation paragraph below.
On the real-recording category, StreamSE obtains the highest BAK.
Overall, StreamSE matches or exceeds the perceptual quality of the offline generative models in our comparison while operating under a fully causal constraint with zero look-ahead.

\vspace{0.2cm}\noindent\textbf{Look-ahead.}
With $\ell$ frames of look-ahead, the input at time $t$ additionally concatenates the noisy latents $n_{\mathrm{embed},t+1},\dots,n_{\mathrm{embed},t+\ell}$; this widens the input projection by $\ell d$ and adds about 0.39M parameters per frame of look-ahead, while the remaining per-frame computation is unchanged and every output is delayed by $\ell$ frames.
One frame of look-ahead improves every metric, with the largest gain under reverberation.
A second frame brings yields smaller additional gains on most metrics, suggesting diminishing returns beyond $\ell{=}2$.

\begin{table}[t]
  \centering
  \caption{Level-truncation ablation on the DNS Challenge test set. $\Delta$OVL and $\Delta$SIM are the changes from $k{=}16$ in OVL averaged over the three categories and in SIM averaged over the two synthetic categories. $^{\dagger}$ marks the setting used in Table~\ref{tab:results}.}
  \label{tab:ktrunc}
  \setlength{\tabcolsep}{3pt}
  \renewcommand{\arraystretch}{1.05}
  \resizebox{\linewidth}{!}{%
    \begin{tabular}{lcccccrr}
    \toprule
    \multirow{3}{*}{Levels} & \multicolumn{3}{c}{OVL $\uparrow$} & \multicolumn{2}{c}{SIM $\uparrow$} & \multirow{3}{*}{$\Delta$OVL} & \multirow{3}{*}{$\Delta$SIM} \\
    \cmidrule(lr){2-4} \cmidrule(lr){5-6}
                      & \makecell{With\\Reverb} & \makecell{Without\\Reverb} & \makecell{Real\\Recordings} & \makecell{With\\Reverb} & \makecell{Without\\Reverb} &          &          \\
    \midrule
    $k{=}16$          &     2.973      &      3.346     &      3.000     &     0.660      &     0.818      & \multicolumn{1}{c}{--} & \multicolumn{1}{c}{--} \\
    $k{=}12$          &     3.001      &      3.364     &      3.017     &     0.667      &     0.810      & $+0.021$ & $-0.001$ \\
    $k{=}8^{\dagger}$ &     3.086      &      3.361     &      3.016     &     0.651      &     0.779      & $+0.048$ & $-0.024$ \\
    $k{=}6$           &     3.127      &      3.345     &      3.026     &     0.643      &     0.749      & $+0.060$ & $-0.043$ \\
    $k{=}4$           &     3.142      &      3.340     &      3.023     &     0.605      &     0.698      & $+0.062$ & $-0.088$ \\
    \bottomrule
  \end{tabular}%
  }
\end{table}

\vspace{0.2cm}\noindent\textbf{RVQ level truncation.}
We describe how we set the number of coarse RVQ levels, which Section~\ref{ssec:experiment_setup} fixes at $k{=}8$.
We sweep $k \in \{4, 6, 8, 12, 16\}$, where $k{=}16$ uses the full RVQ, and train all models on the same data for the same number of epochs.
Table~\ref{tab:ktrunc} reports OVL on the three categories together with SIM on the two synthetic categories, which have clean references.
Truncation acts mainly on the with-reverb category: OVL increases monotonically as $k$ decreases, by 0.169 from $k{=}16$ to $k{=}4$, whereas the spread between the highest and lowest of the five settings is only 0.024 without reverb and 0.026 on real recordings.

This improvement under reverberation is consistent with the hypothesis that the fine levels carry residual reverberation, which motivated level truncation.
SIM moves in the opposite direction, decreasing monotonically as $k$ decreases, by 0.120 from $k{=}16$ to $k{=}4$ without reverb.
The two trends are not symmetric, however: from $k{=}16$ to $k{=}8$, a with-reverb OVL gain of 0.113 costs only 0.039 in without-reverb SIM, whereas from $k{=}8$ to $k{=}4$ a further gain of 0.056 costs 0.081, more than tripling the SIM loss per level.
We choose $k{=}8$, which retains most of the with-reverb gain while stopping short of the steep SIM loss.
This choice also matches AudioDec~\cite{wu2023audiodec}, a streamable 48\,kHz codec whose standard configuration likewise uses eight RVQ levels with 1,024-entry codebooks; StreamSE with $k{=}8$ carries the same number of tokens per frame.

\subsection{On-device Evaluation}\label{ssec:experiment_ondevice}
\begin{figure}[t]
  \centering
  \includegraphics[width=1.01\linewidth]{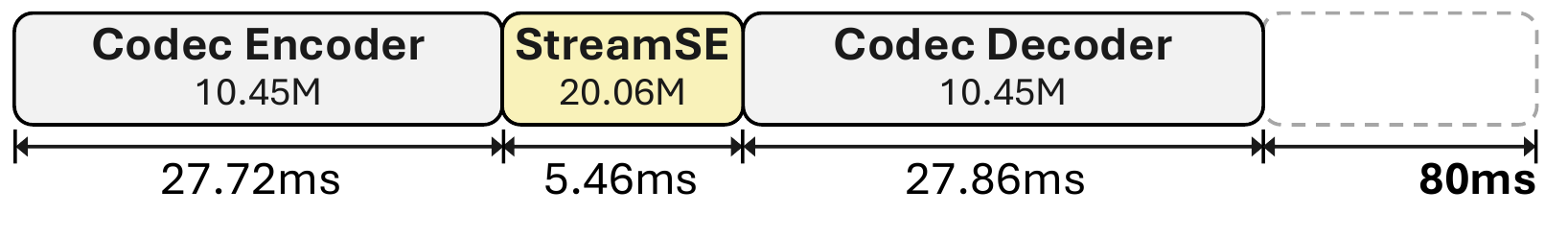}
  \caption{Per-frame runtime breakdown of the streaming pipeline on a Galaxy S22 (single big core, ONNX Runtime CPU backend, dynamic int8 weights, $k{=}8$). Runtime is reported as the p95 per frame and shown against the 80\,ms frame budget.}
  \label{fig:ondevice}
\end{figure}

We evaluate the full pipeline on a Samsung Galaxy S22 (Snapdragon 8~Gen~1), pinned to a single big core on the ONNX Runtime CPU backend, processing one 80\,ms frame at a time with all weight matrix multiplications dynamically quantized to int8.
The int8 row of Table~\ref{tab:results} reports the DNSMOS of this pipeline, whereas the other StreamSE rows use fp32.
The KV cache spans 125 frames, matching the 10\,s training segment, and the cache and feedback state are cleared at the end of each segment.
Each run processes frames back to back for 330\,s under thermal controls (battery power, a temperature gate, and cooling between runs); we report the p95 per-frame time, for the total and each stage independently, after a warm-up interval, averaged over repeated runs whose totals agreed within 0.2\,ms.
The total per-frame time is 61.04\,ms at p95, 76\% of the 80\,ms budget, and no frame exceeds the budget in any run; the encoder, StreamSE, and the decoder require 27.72, 5.46, and 27.86\,ms, respectively, so the codec accounts for 91\% of the latency and StreamSE for 9\% (Fig.~\ref{fig:ondevice}).
The system thus meets the real-time frame budget on a single smartphone CPU core without server-side computation.

\section{Conclusion}\label{sec:conclusion}

We presented StreamSE, a lightweight streaming speech enhancement model that regresses clean continuous latents in the latent space of a frozen neural audio codec with a single causal forward pass per frame and no iterative sampling.
Truncating the RVQ to its coarse levels further improved quality under reverberation while largely preserving speaker identity.
Without look-ahead, StreamSE matched current high-performing offline generative models on objective perceptual metrics, and the complete codec--StreamSE pipeline ran in real time on a single smartphone CPU core with StreamSE accounting for under 10\% of the per-frame latency.
Together, these results establish continuous latent regression on a frozen codec as a practical route to generative enhancement under fully causal, on-device constraints.

\newpage

\section{Compliance with Ethical Standards}
This study used existing publicly available datasets, and no new data were collected. No ethical approval was required.

\bibliographystyle{IEEEbib}
\bibliography{reference}
\end{document}